# Effect of Under-layer Induced Charge Carrier Substitution on the Superconductivity of $Ti_{40}V_{60}$ Alloy Thin Films

Shekhar Chandra Pandey[1,2], Shilpam Sharma[1*], Pooja Gupta[2,3], L. S. Sharath Chandra[1,2] and M. K. Chattopadhyay[1,2]

[1] *Free Electron Laser Utilization Laboratory, Raja Ramanna Center for Advanced Technology, Indore, Madhya Pradesh - 452013, India*

[2] *Homi Bhabha National Institute, Training School Complex, Anushakti Nagar, Mumbai 400 094, India*

[3] *Accelerator Physics and Synchrotrons Utilization Division, Raja Ramanna Center for Advanced Technology, Indore, Madhya Pradesh - 452013, India*

* *shilpam@rrcat.gov.in*

## Abstract

The influence of metallic and semiconducting (V, Al, and Si) under-layer induced charge carrier substitution on the superconducting properties of the $Ti_{40}V_{60}$ alloy thin films are studied and also compared with a pristine reference film without any under-layer. All the films exhibit metallic behavior in the normal state and a superconducting transition at low temperatures, where the superconducting transition temperature is tunable between 4.77 K and 5.73 K. Hall measurements on the films reveal that the under-layer strongly affects the charge carrier type and density, leading to a correlation between increasing carrier concentration and decreasing $T_C$. The Si under-layer introduces the highest disorder, yet yields the highest $T_C$. This indicates that in the $Ti_{40}V_{60}$ alloys, a moderate amount of disorder suppresses the spin-fluctuations (inherent to the alloy system) induced pair breaking, thereby enhancing the superconductivity. The comparable $T_C$ of the film with V under-layer and the film without under-layer, and the much smaller coherence length (~6.2 nm) as compared to the film thickness (25 nm), confirm the absence of any significant proximity effects. These findings demonstrate that under-layer engineering provides an effective route to tune the superconducting properties of Ti-V alloy thin films.

## Introduction

The quest to understand and control superconductivity has been a topic of central attention in condensed matter physics [1]. Since the discovery of superconductivity in mercury, efforts to raise the superconducting transition temperature ($T_C$) and explain the mechanisms behind electron pairing have motivated decades of research across elemental [2-5], alloy [6-9], and unconventional superconductors [10, 11]. Chemical doping [12-14], substitution, application of high-pressure [15, 16] and metallurgical modifications [17-19] are well known methods that can suitably alter the $T_C$ of the materials. In addition to the classical Bardeen-Cooper-Schrieffer (BCS) theory, which attributes superconductivity to electron-phonon coupling, later studies revealed that the superconducting state can be strongly influenced by competing interactions like magnetic fluctuations [20, 21], electronic correlations [22, 23], and structural disorder [24]. In modern materials science, superconductivity is no longer viewed as an isolated property of a pristine crystal but as an emergent phenomenon that can be engineered through microstructure, interfaces, and dimensionality [25, 26]. Thin-film superconductors offer a particularly rich platform for exploring such effects. When materials are confined to the nanometer scale or interfaced with dissimilar layers, their electronic structure, lattice symmetry, carrier density, etc. can be tuned in ways not possible in their bulk form. The properties of the substrate and under-layer materials can significantly influence the thin film growth mode, phase stability, and residual strain, thereby modifying both the normal state and superconducting transport properties [26, 27]. In recent years, such interfacial engineering has led to remarkable findings, from enhanced superconductivity at oxide interfaces [28, 29] to gate-tuned superconductivity [30, 31]. Even in the conventional metallic systems where electron-phonon coupling dominates, the film thickness, strain, and interfacial bonding can significantly alter the density of states and phonon spectra- leading to modification in the superconducting properties [32, 33].

Transition metal alloys represent a particularly interesting class of superconductors in which the electronic structure and lattice can be continuously tuned by varying the composition [20, 23]. Among these materials, superconductivity in the Ti-V alloy system has recently attracted a renewed focus. The Ti-V alloys form over a large composition range and exhibit superconductivity with $T_C$ values depending delicately on the stoichiometry [9]. Vanadium and titanium have $T_C$ values of 5.4 K [23] and 0.4 K [34] respectively. The intermediate alloys, especially near the $Ti_{40}V_{60}$ composition range, exhibit highly enhanced superconducting properties as compared to other compositions [9, 20, 35]. The enhancement, however, is not solely explained by electron-phonon coupling strength or the density of states at the Fermi level. Instead, the Ti-V alloys are known to lie close to the Stoner instability limit [$IN(E_F) \sim 1$] [20], where $I$ is the exchange interaction and $N(E_F)$ is the electronic density of states at the Fermi level. The Ti-V alloys are conventional superconductors, but are strongly influenced by dynamical electronic correlations and the level of disorder [23]. The superconducting properties of Ti-V alloys are significantly influenced by the combined effects of disorder and spin fluctuations. Disorder introduced by Ti substitution modifies the electronic structure and enhances the electron-phonon coupling, which can lead to an increase of $T_C$ [9]. However, spin fluctuations inherent to the Ti-V alloys suppresses the superconductivity by scattering the conduction electrons and thereby hindering the process of formation of the Cooper pairs. The competition between disorder-enhanced electron-phonon interaction and spin-fluctuation-induced pair breaking results in the unique non-monotonic behavior of $T_C$ observed in the Ti-V alloys. The superconductivity in Ti-V alloys thus involves a balance between strong coupling and spin fluctuation effects. Matthias *et al*. studied the variation of $T_C$ of the transition metal superconductors and correlated the superconductivity with the valance electron concentration, i.e., the number of valence electrons per atom ($e/a$) [36]. Alloys with $e/a$ near 4.6- 4.7, e.g. Ti-V, display anomalous deviations from the expected trends of $T_C$ as a function of carrier concentration. In this context, factors that alter the carrier density or local electronic structure can modify the strength of spin fluctuations and hence the $T_C$. Interestingly, moderate disorder can sometimes enhance the $T_C$ by diminishing spin fluctuation effects, a behavior opposite to that expected in the conventional phonon-mediated superconductors [21, 23]. While extensive studies exist on the bulk Ti-V alloys [9, 19-21, 37-41], thin films of Ti-V offer new path for exploring how superconductivity evolves under dimensional confinement and the influence of interfacial properties. Superconductivity in the Ti-V alloy thin films has been studied earlier and our previous studies show that it is possible to tune their $T_C$ by tuning different deposition parameters [35, 42]. The choice of under-layer is especially important, as it governs both the microstructural evolution and the charge carrier environment of the film. Depending on the interfacial energy and lattice matching conditions, different under-layers can also promote the formation and stabilization of distinct crystalline phases in the Ti-V alloys. In the present work, we investigate the influence of under-layer induced carrier substitution and disorder on the superconductivity of $Ti_{40}V_{60}$ alloy thin films. The films are deposited on $SiO_2$ coated Si substrates with three different under-layers: V, Al, and Si. Additionally, a reference film without any under-layer is also studied for comparison. The electrical transport measurements reveal a tunability of $T_C$ from 4.77 K to 5.73 K due to different under-layers. A correlation between the $T_C$ and under-layer induced degree of disorder is presented. These observations point to an interplay between disorder, carrier concentration, and spin fluctuations in the superconductivity of $Ti_{40}V_{60}$ alloy thin films.

## Experimental details

$Ti_{40}V_{60}$ alloy thin films were deposited on 300 nm thick thermal $SiO_2$-coated Si (100) substrates (*p*-type, 0.5 mm thick) using a home-built ultra-high vacuum ($\sim 2 \times 10^{-8}$ mbar) magnetron sputtering system [42]. The alloying of Ti and V was achieved through DC co-sputtering from Ti (99.99%) and V (99.9%) targets in an ultra-pure Ar (99.9995%) environment at ambient temperature. The substrates were mounted on a rotating holder, which rotated at 25 RPM to ensure the homogeneity of the films. The substrate-target distance was kept at 11 cm, and the argon background pressure was maintained at $7 \times 10^{-4}$ mbar. Before deposition, the substrates were cleaned ultrasonically in boiling acetone and ethanol, rinsed with deionized water and blow-dried. To explore the effect of different under-layers on the electronic and superconducting properties of the $Ti_{40}V_{60}$ alloy thin films, V, Al and Si under-layers, each with 10 nm thickness, were deposited on the substrate before the $Ti_{40}V_{60}$ alloy deposition. The alloy film thickness was kept constant at 25 nm. The deposition rates of the individual targets were calibrated beforehand, and the sputtering currents for the Ti and V targets were adjusted to achieve the desired stoichiometry [42, 43]. Four different samples were deposited by varying the under-layers, starting with the $Ti_{40}V_{60}$ alloy thin film deposited on $SiO_2$/Si substrates without any under-layer, followed by samples with the under-layers mentioned above. Sputtering of Ti and V was performed at constant currents of 103 mA and 147 mA, respectively, for all the samples.

Thicknesses and phase purity of the films were measured using X-ray reflectivity (XRR) and grazing incidence X-ray diffraction (GIXRD) measurements respectively, performed using a Bruker, GmbH make D8 diffractometer. The XRR pattern was fitted using REFLEX software [44].

For the in-field transverse and longitudinal transport measurements, the Ti-V alloy films with different under-layers were patterned in the Hall bar geometry using a home-made Direct Laser Writer (DLW) based photolithography setup. The DLW setup utilizes a compound microscope with objective lenses of different magnifications to focus a 405 nm UV diode laser onto a positive photoresist-coated substrate. A minimum feature size of ~5 μm can be achieved using the DLW setup. The laser beam is rastered across the substrate's surface as per the soft mask generated using KLayout (Layout Editor; KLayout 0.28.7) software. After exposure and development process where the photoresist is removed from the exposed areas, the thin film is deposited. The thin film in Hall bar geometry is then retrieved using lift-off technique.

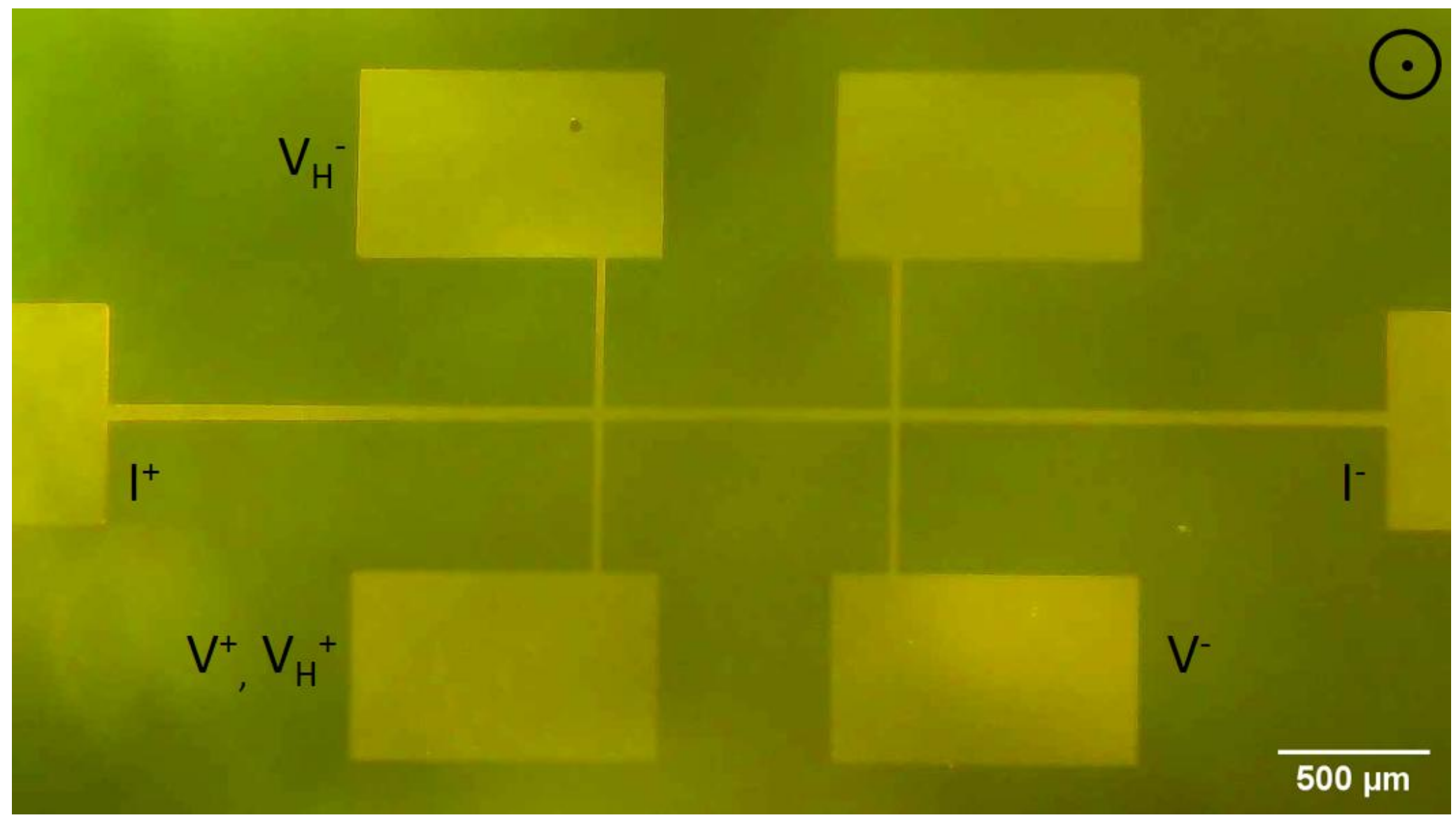

**Figure 1.** Ti-V thin film in the Hall bar geometry synthesized using the Direct Laser Writer based photolithography. The structure was transferred on to positive photoresist coated substrate and the final geometry has been recovered using lift-off process. $I^+$ and $I^-$ are the current contact pads. $V^+$ and $V^-$ are used as voltage contact pads for resistivity measurements, and $V_H^+$ and $V_H^-$ serve as the voltage contact pads for Hall measurements. Image of the Hall bar geometry was taken using an optical microscope.

Temperature and magnetic field dependence of electrical resistivity, and Hall voltage were measured using a 9 T Physical Properties Measurement System with Evercool II Dewar upgrade (PPMS, Quantum Design, USA). Figure 1 shows the Hall bar geometry (width of the bar = 40 $\mu$m). Thin copper wires were used as the electrical leads, and room-temperature curable high-conductivity silver paste was used to prepare the electrical contacts. A constant current was applied across the Hall bar, while the voltages was recorded from the longitudinally arranged contacts for the electrical resistivity measurements and transverse contacts for the Hall voltage measurements. The magnetic field was applied perpendicular to the direction of current flow and normal to the film surface.

## Results and Discussions

The thickness, roughness and density of the alloy films were precisely determined through the analysis of XRR patterns shown in figure 2(a) shows the fitted XRR patterns. All the films exhibit a thickness of approximately 25 nm, with an under-layer thickness of 10 nm. The surface roughness of the deposited $Ti_{40}V_{60}$ films was extracted from the fits and lies in the range 0.5- 1 nm. The mass density obtained from the XRR analysis is approximately 5.3 gm/cm$^3$, consistent with the expected values for $Ti_{40}V_{60}$ alloys. Figure 2 (b) shows the GIXRD patterns of the $Ti_{40}V_{60}$ films deposited on different under-layers. The plots of normalized intensity as a function of diffraction angle are presented in the figure. All the samples exhibit the (110) and (211) reflections of the *bcc* crystal structure with $Im\bar{3}m$ space group, confirming single-phase growth. The peak positions remain nearly unchanged except for the Al under-layer sample. However, the peak shapes and intensities vary with the under-layer, indicating that the interface influences the early stage nucleation and microstructure of the films [45]. The use of an under-layer leads to sharper and more defined diffraction peaks, reflecting improved structural ordering as compared to the film grown directly on the substrate. The lattice parameters of the $Ti_{40}V_{60}$ thin films, determined by fitting the GIXRD data using pseudo-voigt function, were found to be 3.172 Å for the sample without an under-layer, and 3.175 Å, 3.167 Å and 3.204 Å respectively for the Si, V and Al under-layers. The choice of under-layer clearly influences the structural characteristics of the films. Si under-layer causes a slight increase in the lattice parameter relative to the film without under-layer. The V under-layer results in a small decrease and the Al under-layer produces the largest lattice parameter. The results show that different under-layers impact the crystal structure of the deposited $Ti_{40}V_{60}$ thin films.

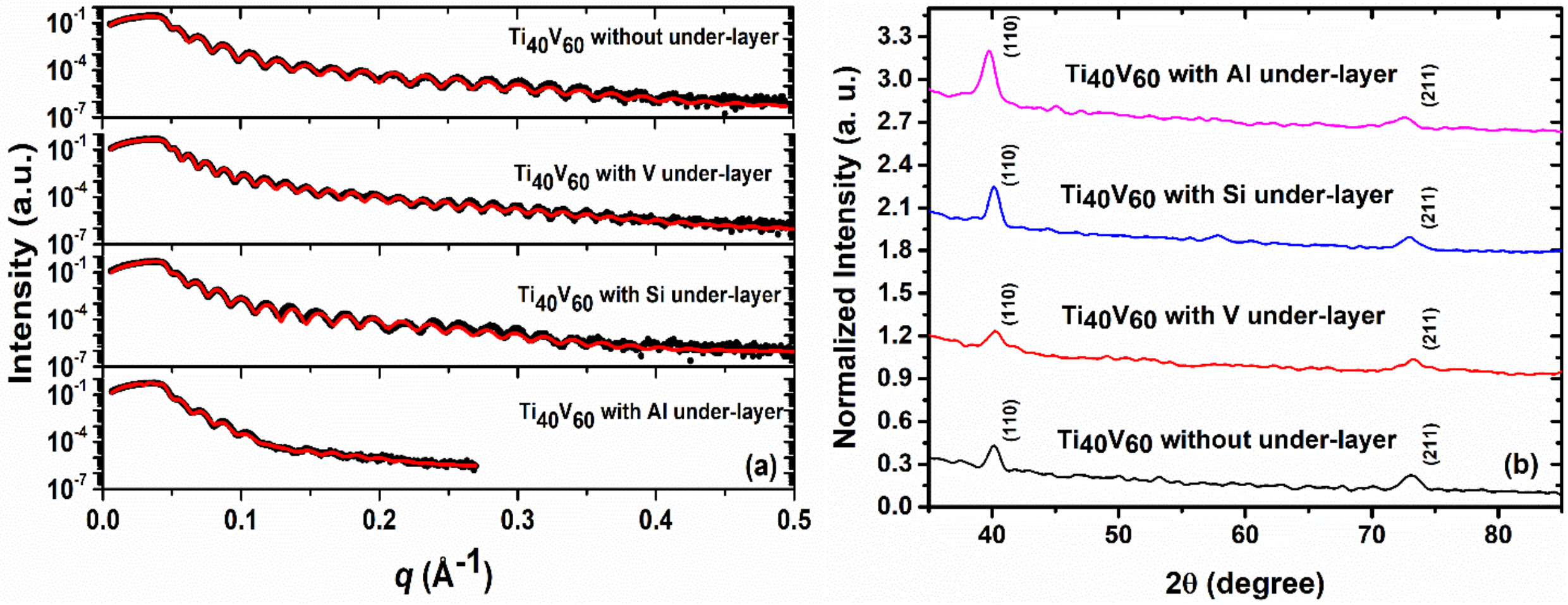

**Figure 2. (a)** XRR patterns as a function of the scattering vector for four $Ti_{40}V_{60}$ thin films with different under-layers. The black curves represent the observed experimental data, while the red curves correspond to the fitted patterns. **(b)** GIXRD patterns of $Ti_{40}V_{60}$ alloy thin films showing the polycrystalline nature of the films.

Figure 3(a) shows the electrical resistivity as a function of temperature for the present films in the range 2- 300 K, and figure 3(b) shows the normalized electrical resistivity in the temperature range 4- 8 K. All the samples display metallic behavior in the normal state and exhibit a transition to zero resistance at low temperatures. The $T_C$, residual resistivity ratios ($RRR$), and sheet resistances ($R_S$) are summarized in Table 1. The $T_C$ is defined as the mean-field transition temperature, corresponding to the point at which the resistance drops to 50% of its normal-state value. By changing the under-layer material, the $T_C$ is tuned from 4.77 K to 5.73 K indicating a clear influence of the underlying layer on the superconducting properties of $Ti_{40}V_{60}$ alloy thin films. The film with a Si under-layer has the highest $T_C$, while the film with an Al under-layer has the lowest $T_C$.

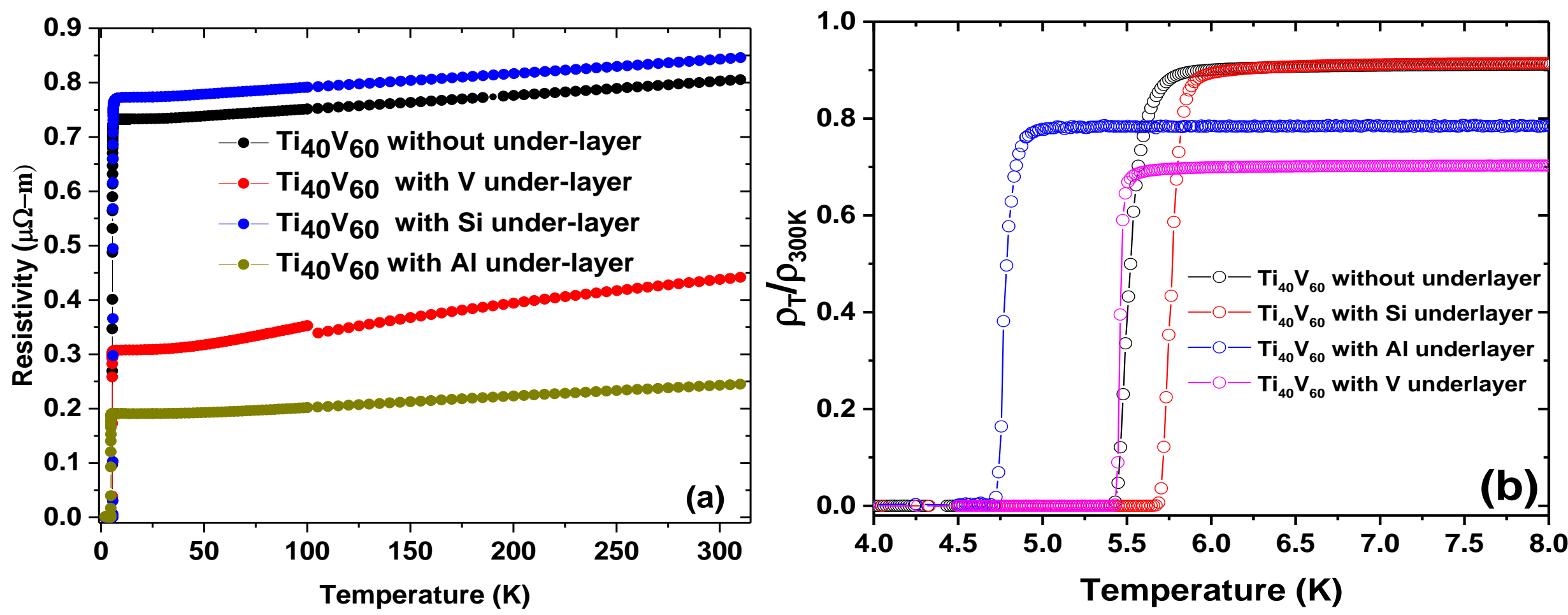

**Figure 3. (a)** Electrical resistivity as a function of temperature for $Ti_{40}V_{60}$ alloy thin films with different under-layers. (b) Normalized electrical resistivity versus temperature curves showing the signature of $T_C$ of the films.

**Table 1.** Comparison of Superconducting transition temperature ($T_C$), residual resistivity ratio ($RRR$) and sheet resistance ($R_S$) of the films

| S. No | Sample | $T_C$ (K) | *RRR* | $R_S$ (Ohms/Square) |
|---|---|---|---|---|
| **1.** | $Ti_{40}V_{60}$ with Si under-layer | 5.73 | 1.09 | 33.7 |
| **2.** | $Ti_{40}V_{60}$ without under-layer | 5.48 | 1.09 | 32.1 |
| **3.** | $Ti_{40}V_{60}$ with V under-layer | 5.46 | 1.42 | 17.5 |
| **4.** | $Ti_{40}V_{60}$ with Al under-layer | 4.77 | 1.28 | 9.72 |

Hall voltage ($V_H$) was measured at different temperatures down to 10 K. Initially, the Hall measurements were performed up to ±5T and no deviation from linear behavior was observed. Therefore, the subsequent measurements at all temperatures were carried out within ±2T magnetic fields. Figure 4(a) shows the $V_H$ as a function of applied magnetic field ($H$) for all the films measured at 300 K. These curves are linear for all the samples. The Hall coefficient ($R_H$) is determined from the slope of the $V_H$ vs. $\mu_0 H$ curve. It can be seen from the figure that except for Al under-layer, all the films have positive slope. This suggest hole-like charge carriers for the films with Si and V under-layers, and for the film with no underlayer, but electron-like charge carriers for the film with Al-under. This observation is consistent with the intrinsic characteristics of the respective under-layer materials. V is known to exhibit predominantly hole-type transport [46], which can influence the overlying film through interfacial charge transfer. In contrast, Al is a free electron metal with electron like charge carriers [47]. Its incorporation as an under-layer can modify the local electronic environment of the film such that electron-type carriers become dominant. The agreement between the measured carrier-type signature and the inherent transport nature of the under-layer materials further supports the interpretation of the observed slopes. To rule out any artefact in the Hall measurements and for confirmation of Al under-layer induced electron doping, the measurements were repeated on samples from two different batches. Both sets of measurements indicated electron-type carriers with identical charge carrier densities for the films with Al underlayer. The type of charge carrier (electron or hole), carrier density ($n$) and carrier mobility ($\mu$) is determined from the Hall measurement and the obtained values are listed in table 2. Since the degree of electronic disorder strongly influences the transport and superconducting properties of the films, the level of disorder in each film was also quantified using the Ioffe–Regel parameter ($K_F l_e$), defined as

$$K_F l_e = \frac{(3\pi^2)^{2/3} \hbar (R_H)^{1/3}}{\rho q^{5/3}} \tag{1}$$

Here, $K_F$ is the Fermi wave vector, $l_e$ is the electronic mean free path, $R_H$ is the Hall coefficient, $\rho$ is the electrical resistivity and $q$ is the electronic charge. The Ioffe-Regel parameter reflects how disorder in a material affects the electron motion. As disorder increases, the mean free path of electrons becomes shorter. When the Ioffe–Regel parameter is close to unity, the average distance an electron travels between scattering events is about the same as the spacing between atoms, meaning that the electrons are strongly scattered, and their movement through the material is hindered, and the material approaches the limit where it may begin to lose its metallic properties. Higher values of $K_F l_e$ implies less disorder in the material.

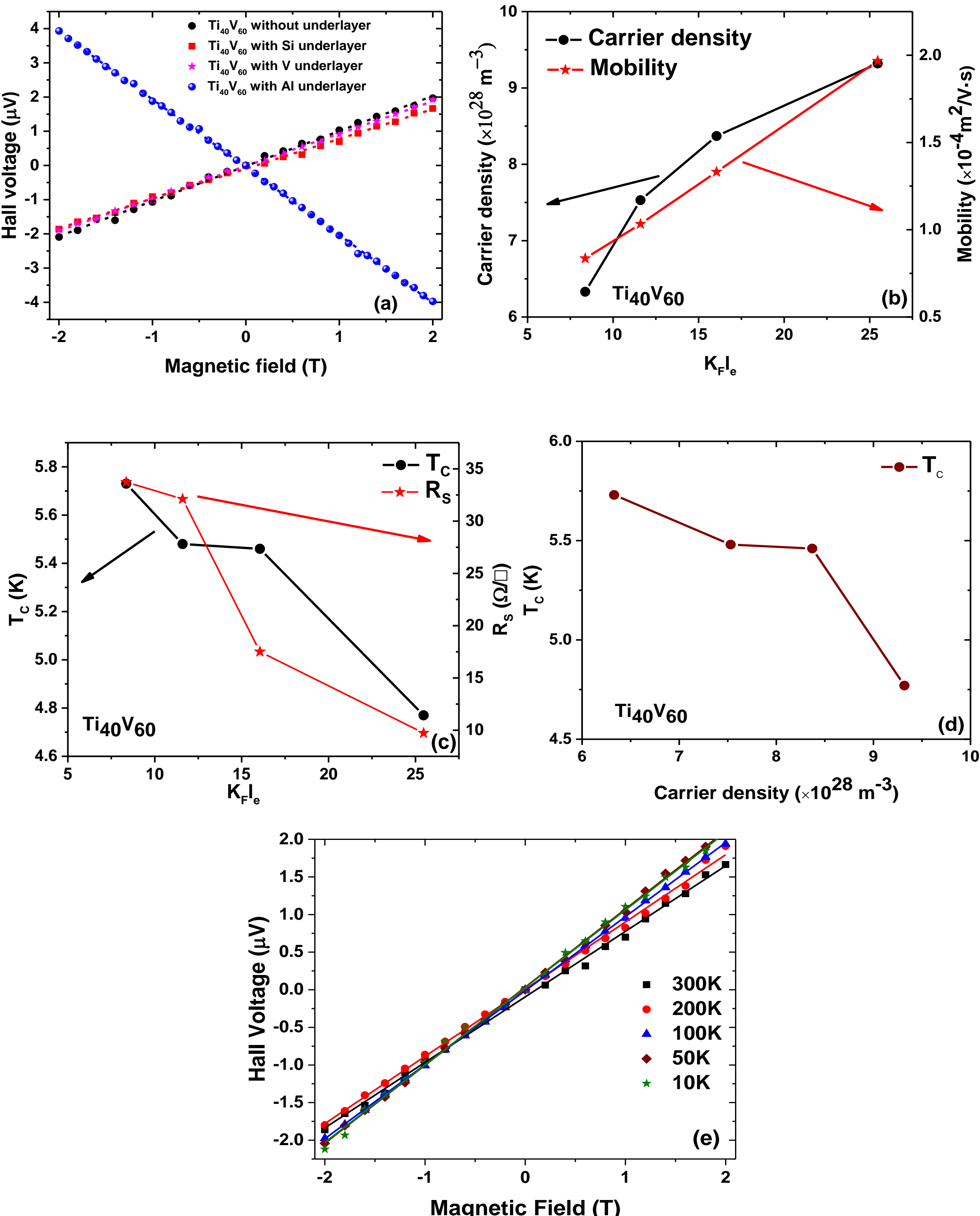

**Figure 4.** **(a)** Hall voltage ($V_H$) as a function of applied magnetic field ($H$) at 300 K, for $Ti_{40}V_{60}$ alloy thin films with different under-layers. Filled dots represents the experimental data and dashed lines indicates the linear fitting of the data. **(b)** Charge carrier density and mobility as a function of disorder for the $Ti_{40}V_{60}$ alloy thin films. **(c)** $T_C$ and $R_S$ of the films as a function of disorder. $T_C$ of the films decreases with increasing disorder. **(d)** $T_C$ of the $Ti_{40}V_{60}$ alloy thin films with carrier density. $T_C$ of the films decreases with increasing charge carrier density. **(e)** Hall Voltage as a function of applied magnetic field at different temperatures for the sample with the highest $T_C$.

**Table 2.** Comparison of carrier type, carrier density (n), carrier mobility ($\mu$) and disorder parameter ($K_Fl_e$) of the four $Ti_{40}V_{60}$ alloy thin films having different under-layers.

| **S. No.** | **Sample** | **Carrier type** | ***n*** **($m^{-3}$)** | ***μ*** **($m^2/V\cdot s$)** | **$K_Fl_e$** |
|---|---|---|---|---|---|
| **1.** | $Ti_{40}V_{60}$ with Si under-layer | Hole | $6.33 \times 10^{28}$ | $8.36 \times 10^{-5}$ | 8.36 |
| **2.** | $Ti_{40}V_{60}$ without under-layer | Hole | $7.53 \times 10^{28}$ | $1.03 \times 10^{-4}$ | 11.60 |
| **3.** | $Ti_{40}V_{60}$ with V under-layer | Hole | $8.37 \times 10^{28}$ | $1.33 \times 10^{-4}$ | 16.05 |
| **4.** | $Ti_{40}V_{60}$ with Al under-layer | Electron | $9.32 \times 10^{28}$ | $1.97 \times 10^{-4}$ | 25.47 |

It can be seen from figure 4(b) that the carrier density and mobility increase with decreasing disorder. In the highly disordered systems (e.g., due to structural defects, impurities, or irregularities), the charge carriers experience enhanced scattering and localization, which reduces their mobility and effective contribution to the conduction. At lower disorder, the charge carriers move more freely. This results in an increase in the effective carrier density, as more carriers are able to participate in conduction. Therefore, a reduction in disorder typically leads to an increase in the measured carrier density. Disorder alters the local atomic environment. Since the *d*-orbitals are sensitive to changes in bond length and coordination, even small distortions modify their overlap [48]. This changes the width of the *d*-band and the density of states near the Fermi level- increasing scattering and altering the alloy's electronic and bonding properties. This effect is consistent with the known influence of disorder in transition metal alloys, where the overlap and interaction of *d*-electrons near the Fermi level dominate the bonding and the electronic properties [23].

Electrical transport studies on the four samples show a clear dependence of the disorder in the films on the choice of under-layer. The Si under-layer introduces the highest degree of disorder in the $Ti_{40}V_{60}$ alloy film. The Al under-layer results in the most ordered film and the V under-layer comes in between, as reflected by the Ioffe-Regel parameter which is lowest for the Si under-layer and highest for the Al under-layer. Interestingly, we find that the highest $T_C$ is for the film grown on Si under-layer, whereas the $T_C$ is lowest for the film with the Al under-layer [figure 4(c)]. The V under-layer gave an intermediate $T_C$, comparable to the film without an under-layer. As the disorder decreases across the different samples, the sheet resistance ($R_S$) of the films systematically decreases, accompanied by a reduction in $T_C$. This relationship suggests that the suppression of disorder enhances carrier mobility, thereby reducing resistive scattering in the normal state. However, the Ti-V alloys are special because they exhibit spin fluctuations that tend to suppress superconductivity [21, 23]. Several experimental and theoretical studies on the Ti-V alloys have shown the presence of spin fluctuations and their influence on superconductivity. Analysis of magnetic susceptibility, heat capacity and resistivity in the Ti-V alloys reveals signatures of itinerant spin fluctuations and indicates that these fluctuations reduce the effective pairing strength unless they are suppressed [20, 21]. First-principles studies of V and V-based alloys further demonstrate that including the spin-fluctuation effects is essential for quantitatively predicting the $T_C$, because spin fluctuations act like a competing (pair-breaking or pair-renormalizing) channel that reduces the $T_C$ relative to the pure electron-phonon mediated superconductivity [23].

Figure 4(d) shows that the $T_C$ of the $Ti_{40}V_{60}$ alloy thin films decreases with increasing charge carrier concentration. This trend is in harmony with the observations made by Matthias *et al.* [24], linking the $T_C$ of the transition metal alloys to the valance electron *e/a* ratio. According to that, the $T_C$ tends to peak near particular *d*-electron counts and decreases as the electronic filling moves away from those optimum values. In this framework, the charge carrier dependence of $T_C$ indicated by the present results acquires a clear physical significance. Films with hole-like carriers (Si, V, and without under-layer) exhibit higher $T_C$ as compared to those with electron-like carriers (Al under-layer film). Hole doping effectively shifts the Fermi level towards lower density of states, thereby reducing the Stoner enhancement factor and weakening the spin fluctuations. The suppression of spin-fluctuation induced pair breaking strengthens the effective electron-phonon coupling, leading to higher $T_C$. Conversely, the electron-like carriers increase the density of states at the Fermi level, enhancing spin fluctuations and reducing $T_C$. In this context, the disorder introduced by the under-layer further helps to weakening these spin fluctuations and resulting to reduce their negative effect on the $T_C$.

The temperature dependence of Hall effect for the present thin film system is depicted through the representative figure in panel 4(e). This figure shows the $V_H$ vs. $\mu_0 H$ curves at different temperatures for the $Ti_{40}V_{60}$ film with Si under-layer, which has the highest $T_C$. The nature of the field dependence of $V_H$ for a particular sample does not change qualitatively as a function of temperature.

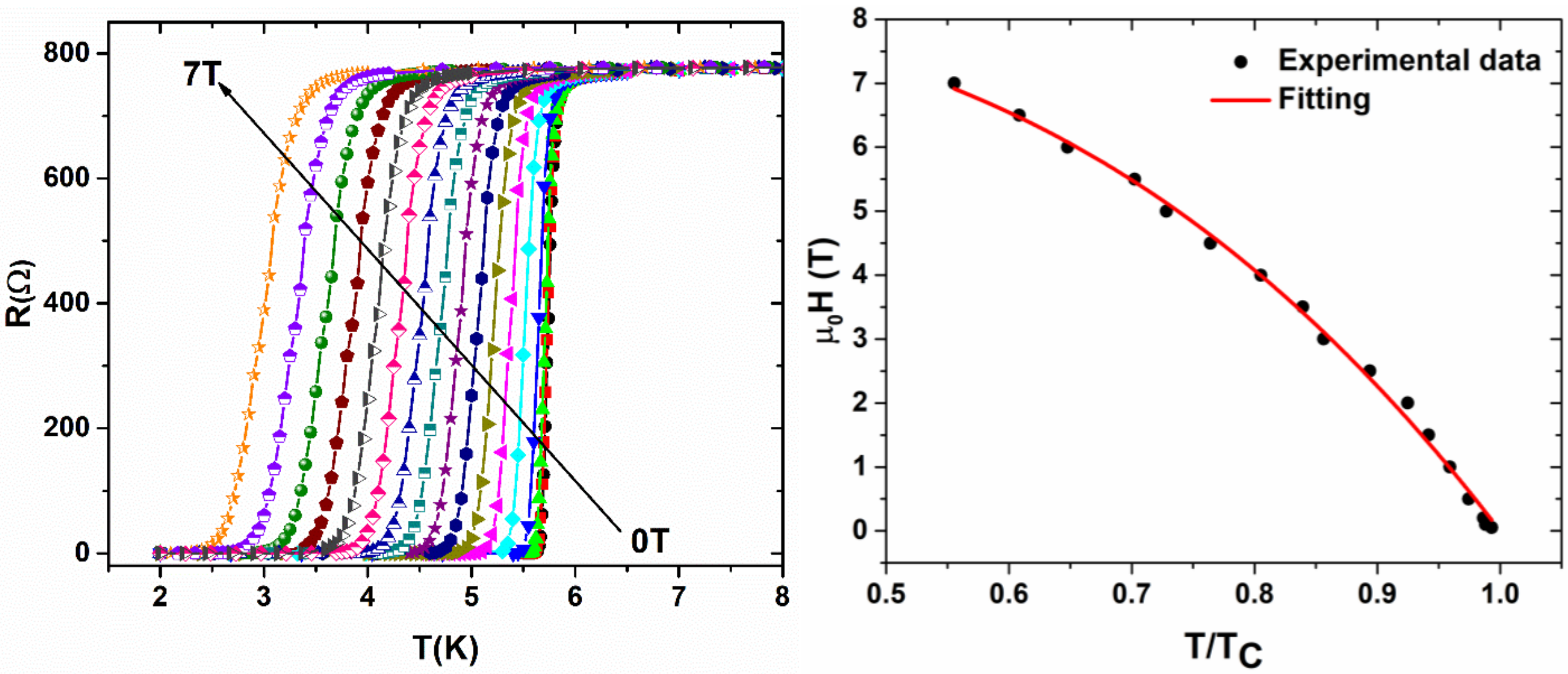

**Figure 5.** **(a)** Resistance as a function of temperature, *R(T)*, for $Ti_{40}V_{60}$ alloy thin film with Si under-layer (highest $T_C$), measured in different applied magnetic fields from 0 to 7 T. **(b)** Critical magnetic field $H_{C2}$ as a function of reduced temperature $T/T_C$; the red curve indicates the fitting of the $H_{C2}$ using equation (2).

The systematic variation of $T_C$ across different under-layers can be understood as an intrinsic effect rather than arising from any proximity-induced influence of the interfacial layer, which could be non-superconducting or superconducting at lower temperatures. This is supported by the observation that the film grown on the V under-layer exhibits a $T_C$ comparable to that of the film deposited directly on the Si substrate, despite the former being in contact with a metallic or superconducting layer. If proximity effects played a dominant role, a significant deviation in $T_C$ would be expected in this case. Additional support for the above argument comes from the estimation of superconducting coherence length ($\xi$). For the $Ti_{40}V_{60}$ thin film having the highest $T_C$ (with Si under-layer), the $\xi$ has been estimated using magnetotransport measurements as shown in figure 5(a). These measurements were performed in fields up to 7 T. Estimation of the upper critical field at 0 K [$H_{C2}(0)$] was done by fitting the $T_C$ versus $\mu_0 H$ curve using equation (2) given below, keeping the $H_{C2}(0)$ as a fitting parameter [Figure 5(b)].

$$H_{C2}(T) = H_{C2}(0)\left[1 - \left(\frac{T}{T_C}\right)^2\right] \quad (2)$$

The fitted value of the $\mu_0 H_{C2}$ at $T = 0$ is found to be 8.3 (±0.1) T. The coherence length estimated using the standard expression,

$$\xi = \sqrt{\frac{\phi_0}{2\pi H_{C2}(0)}} \quad (3),$$

comes out to be 6.2 nm. The film thickness, on the other hand, is 25 nm. Since the film thickness is nearly four times larger than $\xi$, any proximity-induced suppression of the superconducting order parameter at the interface is confined to a thin region near the under-layer and does not influence the bulk of the film. Hence we conjecture that the observed difference in $T_C$ arises from modifications in the film's intrinsic properties, such as charge carrier density rather than any interfacial or under-layer induced proximity effect. The crystal structure and metallurgical phase-stabilizing nature of the under-layers could also play a crucial role in the electronic properties of the material. Si and V are $\beta$-phase stabilizers [49], having a diamond cubic and a body-centered cubic (*bcc*) structure respectively. Both Si and V promote the $\beta$-phase in $Ti_{40}V_{60}$ alloy thin films, and thus favours higher $T_C$. In contrast, Al is a $\alpha$-stabilizer with a face-centered cubic (*fcc*) structure that suppresses the $\beta$-phase [49, 50], leading to lower $T_C$. Changes in peak positions of the GIXRD patterns of the present films [figure 2(b)] show how the $Ti_{40}V_{60}$ films structurally responds to different under-layers. The $\beta$-stabilizing layers like Si and V shift the lattice toward the $\beta$-phase, while an $\alpha$-stabilizer like Al suppresses it. Since the $\beta$-phase favours higher $T_C$ in the Ti-V alloys [20], the GIXRD signatures of films grown on Si and V correlate with the higher $T_C$. The film on Al shows GIXRD features consistent with α-phase stabilization, which correlates with its lower $T_C$. Furthermore, the fact that the Si under-layer, in spite of being amorphous, yields the highest $T_C$, strongly indicates that the observed variation in $T_C$ cannot be attributed merely to structural epitaxy or lattice matching with the under-layer. Instead, it highlights that the under-layer primarily modifies the intrinsic disorder and carrier characteristics of the $Ti_{40}V_{60}$ layer rather than imposing structural or proximity-driven effects.

## Conclusions

In conclusion, we have demonstrated that the $T_C$ of $Ti_{40}V_{60}$ alloy thin films can be effectively tuned by varying the under-layer material without changing the alloy composition. The observed variation of $T_C$ from 4.77 K to 5.73 K arises from under-layer induced modulation of carrier density and disorder. The Si under-layer, which introduces the highest disorder, yields the highest $T_C$, while the Al under-layer produces the lowest. This highlights the role of spin fluctuations in the Ti-V alloys, where moderate disorder suppresses these pair-breaking excitations, thereby enhancing the superconductivity. The correlation between decreasing carrier density and increasing $T_C$ further supports a spin fluctuation mediated mechanism. The comparable $T_C$ values in the films with V under-layer and without any under-layer, together with the short coherence length (6.2 nm) relative to the film thickness (25 nm), confirm that proximity effects are not significant enough in this case. These results establish under-layer engineering as an effective approach to control superconductivity in the Ti-V thin films through charge carrier substitution.

## References

[1] Webb G W, Marsiglio F and Hirsch J E 2015 Superconductivity in the elements, alloys and simple compounds *Physica C: Superconductivity and its Applications* **514** 17

[2] Mito M, Matsui H, Tsuruta K, Yamaguchi T, Nakamura K, Deguchi H, Shirakawa N, Adachi H, Yamasaki T, Iwaoka H, Ikoma Y and Horita Z 2016 Large enhancement of superconducting transition temperature in single-element superconducting rhenium by shear strain *Scientific Reports* **6** 36337

[3] Cao Z-Y, Jang H, Choi S, Kim J, Kim S, Zhang J-B, Sharbirin A S, Kim J and Park T 2023 Spectroscopic evidence for the superconductivity of elemental metal Y under pressure *NPG Asia Materials* **15** 5

[4] Mito M, Tsuji H, Tajiri T, Nakamura K, Tang Y and Horita Z 2024 Superconductivity of barium with highest transition temperatures in metallic materials at ambient pressure *Scientific Reports* **14** 965

[5] Zhang C, He X, Liu C, Li Z, Lu K, Zhang S, Feng S, Wang X, Peng Y, Long Y, Yu R, Wang L, Prakapenka V, Chariton S, Li Q, Liu H, Chen C and Jin C 2022 Record high Tc element superconductivity achieved in titanium *Nature Communications* **13** 5411

[6] Compton V B, Corenzwit E, Maita J P, Matthias B T and Morin F J 1961 Superconductivity of Technetium Alloys and Compounds *Physical Review* **123** 1567

[7] Matthias B T, Geballe T H, Compton V B, Corenzwit E and Hull G W 1962 Superconductivity of Chromium Alloys *Physical Review* **128** 588

[8] Tai M, Inoue K, Kikuchi A, Takeuchi T, Kiyoshi T and Hishinuma Y 2007 Superconducting Properties of V-Ti Alloys *IEEE Transactions on Applied Superconductivity* **17** 2542

[9] Matin M, Sharath Chandra L S, Chattopadhyay M K, Meena R K, Kaul R, Singh M N, Sinha A K and Roy S B 2015 Critical current and flux pinning properties of the superconducting Ti–V alloys *Physica C: Superconductivity and its Applications* **512** 32

[10] Si Q and Hussey N E 2023 Iron-based superconductors: Teenage, complex, challenging *Physics Today* **76** 34

[11] Stewart G R 2011 Superconductivity in iron compounds *Reviews of Modern Physics* **83** 1589

[12] Disiena M N, Jha R, Sloan L and Banerjee S K 2025 Effects of Mo Doping on the Superconducting Properties of NbSe2 *ACS Applied Electronic Materials* **7** 9526

[13] Bohnenstiehl S D, Susner M A, Yang Y, Collings E W, Sumption M D, Rindfleisch M A and Boone R 2011 Carbon doping of MgB2 by toluene and malic-acid-in-toluene *Physica C: Superconductivity* **471** 108

[14] Han D, Ming W, Xu H, Chen S, Sun D and Du M-H 2019 Chemical Trend of Transition-Metal Doping in WSe2 *Physical Review Applied* **12** 034038

[15] Drozdov A P, Eremets M I, Troyan I A, Ksenofontov V and Shylin S I 2015 Conventional superconductivity at 203 kelvin at high pressures in the sulfur hydride system *Nature* **525** 73

[16] Lorenz B and Chu C W 2005 *Frontiers in Superconducting Materials,* ed A V Narlikar (Berlin, Heidelberg: Springer Berlin Heidelberg) pp 459

[17] Ramjan S K, Khandelwal A, Paul S, Chandra L S S, Singh R, Venkatesh R, Kumar K, Rawat R, Dutt S, Sagdeo A, Ganesh P and Chattopadhyay M K 2024 Enhancement of irreversibility field and critical current density of rare earth containing V0.60Ti0.40 alloy superconductor by cold-working and annealing *Journal of Alloys and Compounds* **976** 173321

[18] Ramjan S K, Sharath Chandra L S, Singh R and Chattopadhyay M K 2023 *Proceedings of the 29th International Conference on Low Temperature Physics (LT29)*: Journal of the Physical Society of Japan)

[19] Ramjan S K, Sharath Chandra L S, Singh R, Ganesh P, Sagdeo A and Chattopadhyay M K 2022 Enhancement of functional properties of V0.6Ti0.4 alloy superconductor by the addition of yttrium *Journal of Applied Physics* **131** 063901

[20] Matin M, Sharath Chandra L S, Pandey S K, Chattopadhyay M K and Roy S B 2014 The influence of electron-phonon coupling and spin fluctuations on the superconductivity of the Ti-V alloys *The European Physical Journal B* **87** 131

[21] Matin M, Sharath Chandra L S, Meena R, Chattopadhyay M K, Sinha A K, Singh M N and Roy S B 2014 Spin-fluctuations in Ti0.6V0.4 alloy and its influence on the superconductivity *Physica B: Condensed Matter* **436** 20

[22] Mitra M, Ghosh H and Behera S N 1998 Effect of electron correlation on superconducting pairing symmetry *The European Physical Journal B - Condensed Matter and Complex Systems* **2** 371

[23] Jones D, Östlin A, Weh A, Beiuşeanu F, Eckern U, Vitos L and Chioncel L 2024 Superconducting transition temperatures of pure vanadium and vanadium-titanium alloys in the presence of dynamical electronic correlations *Physical Review B* **109** 165107

[24] Neverov V D, Lukyanov A E, Krasavin A V, Vagov A, Lvov B G and Croitoru M D 2024 Exploring disorder correlations in superconducting systems: spectroscopic insights and matrix element effects *Beilstein journal of nanotechnology* **15** 199

[25] Guo Y, Qiu D, Shao M, Song J, Wang Y, Xu M, Yang C, Li P, Liu H and Xiong J 2023 Modulations in Superconductors: Probes of Underlying Physics **35** 2209457

[26] Maggiora J, Wang X and Zheng R 2024 Superconductivity and interfaces *Physics Reports* **1076** 1

[27] Liu Y, Meng Q, Mahmoudi P, Wang Z, Zhang J, Yang J, Li W, Wang D, Li Z, Sorrell C C and Li S 2024 Advancing Superconductivity with Interface Engineering *Advanced Materials* **36** 2405009

[28] Al-Tawhid A H, Poage S J, Salmani-Rezaie S, Gonzalez A, Chikara S, Muller D A, Kumah D P, Gastiasoro M N, Lorenzana J and Ahadi K 2023 Enhanced Critical Field of Superconductivity at an Oxide Interface *Nano Letters* **23** 6944

[29] Yang B, Zhao C, Xia B, Ma H, Chen H, Cai J, Yang H, Liu X, Liu L, Guan D, Wang S, Li Y, Liu C, Zheng H and Jia J 2023 Interface-enhanced superconductivity in monolayer 1T′-MoTe2 on SrTiO3(001) *Quantum Frontiers* **2** 9

[30] Qiao W, Ma Y, Yan J, Xing W, Yao Y, Cai R, Li B, Xiong R, Xie X C, Lin X and Han W 2021 Gate tunability of the superconducting state at the EuOKTaO3 (111) interface *Physical Review B* **104** 184505

[31] Piatti E 2021 Ionic gating in metallic superconductors: A brief review *Nano Express* **2** 024003

[32] Liu L, Miao G, Liu B, Nan P, Wang Y, Ge B, Zhu X, Yang F, Wang W and Guo J 2019 Interfacial effects on the superconducting properties of LaSi2 (112) films on Si(111) *Physical Review B* **100** 165308

[33] Liu Y, Meng Q, Mahmoudi P, Wang Z, Zhang J, Yang J, Li W, Wang D, Li Z, Sorrell C C and Li S 2024 Advancing Superconductivity with Interface Engineering **36** 2405009

[34] Mito M, Mokutani N, Tsuji H, Tang Y, Matsumoto K, Murayama M and Horita Z 2022 Achieving superconductivity with higher Tc in lightweight Al–Ti–Mg alloys: Prediction using machine learning and synthesis via high-pressure torsion process *Journal of Applied Physics* **131** 105903

[35] Pandey S C, Sharma S, Khandelwal A and Chattopadhyay M K 2025 Ambient temperature growth and superconducting properties of Ti-V alloy thin films *AIP Conference Proceedings* **3198** 020113

[36] Anderson P W and Matthias B T 1964 Superconductivity *Science* **144** 373

[37] Matin M, Chattopadhyay M K, Chandra L S S and Roy S B 2016 High-field paramagnetic Meissner effect and flux creep in low-Tc Ti–V alloy superconductors *Superconductor Science and Technology* **29** 025003

[38] Matin M, Sharath Chandra L S, Chattopadhyay M K, Meena R K, Kaul R, Singh M N, Sinha A K and Roy S B 2013 Magnetic irreversibility and pinning force density in the Ti-V alloys *Journal of Applied Physics* **113** 163903

[39] Matin M, Sharath Chandra L S, Chattopadhyay M K, Singh M N, Sinha A K and Roy S B 2013 High field paramagnetic effect in the superconducting state of Ti0.8V0.2 alloy *Superconductor Science and Technology* **26** 115005

[40] Collings E W 1974 Anomalous electrical resistivity, bcc phase stability, and superconductivity in titanium-vanadium alloys *Physical Review B* **9** 3989

[41] Paul S, Ramjan S, Venkatesh R, Chandra L S S and Chattopadhyay M K 2021 Grain Refinement and Enhancement of Critical Current Density in the V0.60Ti0.40 Alloy Superconductor With Gd Addition *IEEE Transactions on Applied Superconductivity* **31** 1

[42] Pandey S C, Sharma S, Pandey K K, Gupta P, Rai S, Singh R and Chattopadhyay M K 2025 Sputtering current driven growth and transport characteristics of superconducting Ti40V60 alloy thin films *Journal of Applied Physics* **137** 113902

[43] Pandey S C, Sharma S and Chattopadhyay M K 2025 Growth Optimization of MoSi Thin Film and Measurement of Transport Critical Current Density of its Meander Structure *physica status solidi (b)* e202500289

[44] Vignaud G and Gibaud A 2019 REFLEX: a program for the analysis of specular X-ray and neutron reflectivity data *Journal of Applied Crystallography* **52** 201

[45] Spaepen F 2000 Interfaces and stresses in thin films *Acta Materialia* **48** 31

[46] Kuvandikov O K, Hamrayev N S, Rajabov R M and Hamrayeva N N 2023 Anisotropic scattering of charge carriers in double continuous solid solutions of the niobium-vanadium system *Journal of Physics: Conference Series* **2573** 012022

[47] Bradley J M and Stringer J 1974 Hall effect measurements in aluminium alloys *Journal of Physics F: Metal Physics* **4** 839

[48] Huang L-F, Grabowski B, McEniry E, Trinkle D R and Neugebauer J 2015 Importance of coordination number and bond length in titanium revealed by electronic structure investigations **252** 1907

[49] Donachie M J 2000 *Titanium: A Technical Guide, 2nd Edition*: ASM International)

[50] Ballor J, Shawon A A, Zevalkink A, Sunaoshi T, Misture S and Boehlert C J 2023 The effects of Fe and Al on the phase transformations and mechanical behavior of β-Ti alloy Ti-11at.%Cr *Materials Science and Engineering: A* **886** 145677